\documentclass[twocolumn,english,amsmath,amssymb,superscriptaddress,nofootinbib]{revtex4-1}
\usepackage[T1]{fontenc}
\usepackage[latin9]{inputenc}
\usepackage{amsmath}
\usepackage{amssymb}
\usepackage{graphicx}

\makeatletter
\usepackage{babel}
\usepackage{babel}
\usepackage{newtxtext}
\usepackage{xr}
\usepackage{bbold}

\usepackage{color}

\usepackage{ulem}

\newcommand{\stkout}[1]{\ifmmode\text{\sout{\ensuremath{#1}}}\else\sout{#1}\fi}

\makeatother

\usepackage{babel}
\begin{document}
\title{Exact Numerical Solution of Stochastic Master Equations for 
Conditional Spin Squeezing}

\author{ZhiQing Zhang}
\address{Henan Key Laboratory of Diamond Optoelectronic Materials and Devices, Key Laboratory of Material Physics Ministry of Education, School of Physics and Microelectronics, Zhengzhou University, Zhengzhou 450052, China}

\author{Yuan Zhang}
\email{yzhuaudipc@zzu.edu.cn}
\address{Henan Key Laboratory of Diamond Optoelectronic Materials and Devices, Key Laboratory of Material Physics Ministry of Education, School of Physics and Microelectronics, Zhengzhou University, Zhengzhou 450052, China}
\address{Institute of Quantum Materials and Physics, Henan Academy of Sciences, Zhengzhou 450046, China}

\author{HaiZhong Guo}
\address{Henan Key Laboratory of Diamond Optoelectronic Materials and Devices, Key Laboratory of Material Physics Ministry of Education, School of Physics and Microelectronics, Zhengzhou University, Zhengzhou 450052, China}
\address{Institute of Quantum Materials and Physics, Henan Academy of Sciences, Zhengzhou 450046, China}

\author{Chongxin Shan}
\address{Henan Key Laboratory of Diamond Optoelectronic Materials and Devices, Key Laboratory of Material Physics Ministry of Education, School of Physics and Microelectronics, Zhengzhou University, Zhengzhou 450052, China}
\address{Institute of Quantum Materials and Physics, Henan Academy of Sciences, Zhengzhou 450046, China}

\author{Gang Chen}
\email{chengang971@163.com}
\address{Henan Key Laboratory of Diamond Optoelectronic Materials and Devices, Key Laboratory of Material Physics Ministry of Education, School of Physics and Microelectronics, Zhengzhou University, Zhengzhou 450052, China}
\address{Institute of Quantum Materials and Physics, Henan Academy of Sciences, Zhengzhou 450046, China}

\author{Klaus M{\o}lmer}
\email{klaus.molmer@nbi.ku.dk}
\address{Niels Bohr Institute, University of Copenhagen, 2100 Copenhagen, Denmark}

\begin{abstract}
Stochastic master equations are often used to describe  conditional spin squeezing of  atomic ensemble, but are limited so far to the systems with few atoms due to the exponentially increased  Hilbert space. In this article, we present an exact numerical solution of these equations for systems with identical atoms by mapping identical density matrix elements to a single quantity characterized by collective quantum numbers, and apply it to the system with hundred atoms in a bad cavity subject to a homodyne detection. We demonstrate that the spin squeezing can be vividly illustrated by the Gaussian-like  distribution of the collective density matrix elements, and we examine the influence of the probe field strength and polarization, the detection efficiency, the spontaneous emission rate and the number of atoms.  Our exact approach can play an important role in gauging the approximate approaches applied for systems with more atoms, such as Gaussian-state formalism and stochastic mean-field approach, and it permits also exploration of entanglement effects beyond these approaches. 
\end{abstract}
\maketitle

\section{Introduction}

Atomic spin squeezing   has been explored extensively in recent years \cite{JMa,HMWiseman}, because it is  a means to explore quantum entanglement of many particles \cite{OGuhne}, and it holds promise for application in quantum information \cite{MANielsen} and high-precision spectroscopy \cite{DJWineland,VMeyer,ALChauvet}.  Among  various methods to generate spin squeezing \cite{JHald,AS=0000F8rensen},  quantum non-demolition detection (QND) \cite{AKuzmich,ZChen,KCCox} has attracted significant attention, where  measurements accumulate information about an observable that commutes with the system Hamiltonian. Thus, conditioned on the measurement outcome,  the uncertainty of the observable is reduced, leading to the so-called conditional spin squeezing \cite{AKuzmichNP}.  So far, QND spin squeezing has been realized with  atoms in  free space \cite{AKuzmich,JAppel}, atoms coupled weakly \citep{OHosten,MHSchleierSmith} or strongly with  optical cavities \cite{ZChen,KCCox} subject to homodyne or heterodyne detection.  In the former two cases,   Raman scattering of three- or four-level  atoms  is usually explored to map the collective population of the atomic hyper-fine levels to the light phase shift. In the latter case, Rabi splitting due to atoms-photon dressed states is often adopted to infer the collective population.

In the study of the atomic spin squeezing, one often introduces the so-called collective spin vectors $\hat{\bf J} = \sum_{i=x,y,z} \hat{J}_i {\bf e}_i$ with  three components $\hat{J} = (1/2)\sum_{k=1}^N \hat{\sigma}_{k,i}$ in the Cartesian coordinate systems (with ${\bf e}_i$ as the unit vectors). Here, $\hat{\sigma}_{k,i}$ are the three Pauli operators of the $k$-th atom of the total $N$  atoms. By eliminating the higher excited levels in the Raman scattering or the atom-photon dressed states, one can achieve an effective  stochastic master equation (SME) \cite{LKThomsen}, $\partial_t \hat{\rho} = -M  \mathcal{D}[\hat{J}_z]\hat{\rho}  + \sqrt{M} (dW/dt)
 \mathcal{H}[\hat{J}_z]\hat{\rho} $,  to capture the key physics of the measurements,  where  the first and second term on the right side describe  the collective dephasing and the measurement backaction  with a strength $M$.  Here, the  super-operators are defined as  $\mathcal{D}[\hat{o}]\hat{\rho} = \left(\hat{o}^{\dagger}\hat{o}\hat{\rho}+\hat{\rho} \hat{o}^{\dagger}\hat{o}\right)/2-\hat{o}\hat{\rho} \hat{o}^{\dagger}$  and $\mathcal{H}[\hat{o}]\hat{\rho} = \hat{o}\hat{\rho} + \hat{\rho}\hat{o}^\dagger - (\langle \hat{o}\rangle + \langle \hat{o}^\dagger\rangle)  \hat{\rho}$ for any operator $\hat{o}$ and its expectation value $\langle \hat{o}\rangle = \mathrm{tr}\left\{ \hat{o} \hat{\rho}\right\}$.  The random numbers $dW$ account for shot noise in the homodyne or heterodyne detection,  and follow a normal distribution with zero mean and variance $dt$, i.e. $\mathrm{E}\left(dW\right)=0$ and $dW^2=dt$, and the photo-current in these detections is given by $I(t)=2 \sqrt{M}\langle \hat{J}_{z}\rangle(t) +dW(t)/dt$.

 The standard approach to solve the SME is the density matrix approach, where the product states  $\left|\alpha\right\rangle =\prod_{k=1}^N\left|a_{k}\right\rangle $ and $\left|\beta\right\rangle =\prod_{k=1}^N\left|b_{k}\right\rangle $ are defined with the upper and lower   levels $a_k,b_k =\uparrow_k, \downarrow_k$ of the atomic hyper-fine levels, and the  matrix elements $\rho_{\beta\alpha}=\mathrm{tr}\left\{ \left|\alpha \right\rangle \left\langle \beta\right| \hat{\rho}\right\}$ are defined with the transition operators $\left|\beta\right\rangle \left\langle \alpha\right|$. Because  the number of matrix elements $4^N$ scales exponentially with the number of atoms $N$, the density matrix approach can only be used to study systems with few atoms. To overcome this problem, we can assume that all the atoms are identical, and then compared the matrix elements with identical values. To remove redundancy, we can map the identical elements $\{\rho_{\alpha\beta}\}$ to one single quantity denoted by the symbol  $\left\langle n\right\rangle \equiv\langle \begin{array}{cc}
n_{\uparrow \uparrow} & n_{\uparrow \downarrow}\\
n_{\downarrow \uparrow} & n_{\downarrow \downarrow}
\end{array}\rangle$, where the collective numbers $n_{ab}$ count the occurrences of $\left|a_{k}=a\right\rangle $ and $\left|b_{k}=b\right\rangle $ in the states $|\alpha\rangle $ and $\left|\beta\right\rangle$~\citep{MRichter,YZhang2015,YZhang}.  Here, the four numbers $n_{ab}$ are  integers  between $0$ and $N$, and satisfy the conditions $\sum_{a,b}n_{ab}=N$. The number of the sets $\{n_{ab}\}$ satisfying these conditions is given by the binomial function  $C_{N+3}^{3} \approx N^3$.  Since the number of independent quantities scales only polynomially,  we can easily handle systems with hundreds of atoms. Such an exact approach was applied previously by us~\citep{YZhang2015,YZhang} and M. Richter et al.~\citep{MRichter} to study plasmonic lasing, and here we extend it to the field of quantum measurement and control. Furthermore, to efficiently solve the equations for $\left\langle n\right\rangle$, we have also designed an exquisite program, which takes the advantages of GPU parallel computing.

\begin{figure}
\begin{centering}
\includegraphics[scale=0.65]{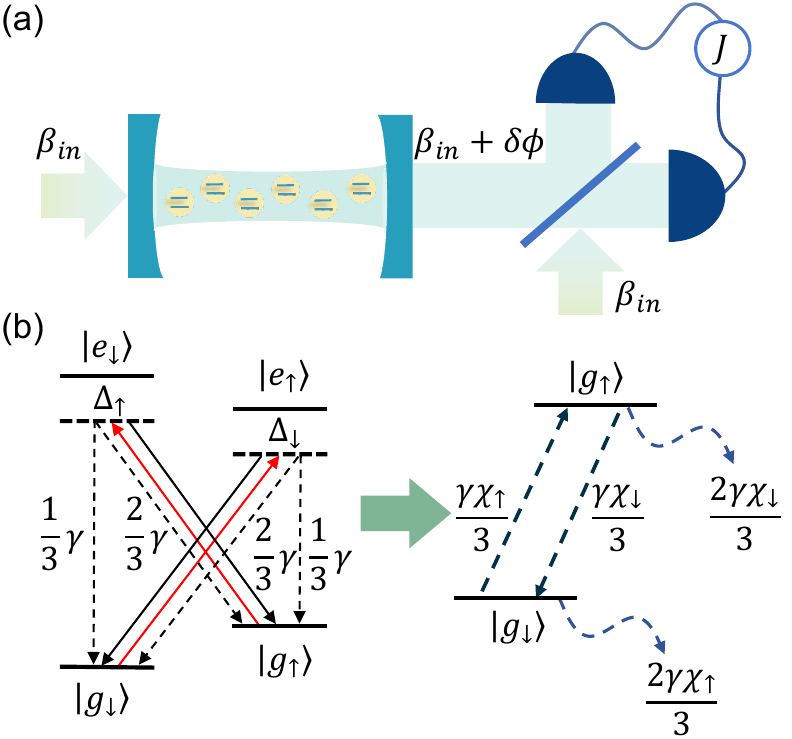}
\par\end{centering}
\caption{\label{fig:system-theory} System and energy diagram. Panel (a) shows that  a probe field with an amplitude $\beta_{in}$ interacts off-resonantly with  atoms in a bad-cavity, and picks up an extra  phase $\delta\phi$ which is proportional to the collective population of two atomic hyper-fine ground levels. The left part of panel (b) shows that  the atoms with two ground and excited hyper-fine levels are driven off-resonantly by circular polarized probe fields, and they experience decay  through vertical and circular transitions (dashed lines), where  $\Delta_{\uparrow}$ and $\Delta_{\downarrow}$ are frequency detuning between the cavity mode  and the atomic transitions. The right part of panel (b) shows the effective diagram after adiabatic elimination of the hyper-fine excited states, where the resulting effective two-level system experiences Raman-induced dephasing, decay, pumping of individual atoms (dashed arrows), collective dephasing and measurement backaction.  }
\end{figure}

In this article, we apply the above technique to study an exemplary system  as considered in Ref. \cite{ASSorensen} [Fig. \ref{fig:system-theory} (a)], where a laser field probes a bad optical cavity,  the cavity interacts with many atoms, and the transmitted field is measured by a balanced homodyne detection. Since the atoms couple with the cavity  off-resonantly,  we can adiabatically eliminate two hyper-fine excited states and the scattered cavity photons to achieve an effective model for two hyper-fine ground states, where the atoms experience the Raman-induced collective dephasing and the QND measurement backaction as elaborated in above. Besides, combined with the spontaneous emission of the excited states, the off-resonant driving leads also to Raman-induced dephasing, decay and pumping of the individual atoms [Fig. \ref{fig:system-theory} (b)].  We simulate such a system with up to hundred atoms, demonstrate the spin squeezing visually by the Gaussian-like  distribution  of the collective density matrix elements, and study the influence of the probe field polarization and amplitude,  the detection efficiency,  the atomic spontaneous emission, as well as the number of atoms.

The current article is organized as follows. In Sec. \ref{sec:esme} we present the stochastic master equation, the derived equations for the collective density matrix elements and the calculation of spin squeezing parameters. In
Sec. \ref{sec:results} we specify the simulation parameters and discuss the numerical results. In  the end,  we conclude our work and discuss possible extensions in future. 

\section{Effective Stochastic Master Equation \label{sec:esme} }

In the Appendix~\ref{sec:sme}, we have presented the stochastic master equation to describe the conditional dynamics of the system in the presence of homodyne detection, as shown in Fig.~\ref{fig:system-theory}(a). There, a probe field with an amplitude $\beta_{in}$ drives off-resonantly four-level atoms in an optical cavity to acquire a phase shift $\delta \phi$, and the mixing of the transmitted and input field is measured by two photodetectors forming a homodyne detection. The atoms are driven by the probe field  through two circular polarized transitions with the frequency detunings $\Delta_\downarrow,\Delta_\uparrow$, they couple with two optical cavity modes with orthogonal polarizations through the same transitions with strength $g$, and they experience simultaneously the excited-state decay with the rate $\gamma$ and the branching ratios as indicated in Fig.~\ref{fig:system-theory}(b).

Under the large frequency detuning conditions, the excited states are barely populated  and the scattered photons are few. After  adiabatically eliminating these quantities, we get the following effective stochastic master equation: 
\begin{align}
\partial_t\hat{\rho} & =-\frac{i}{\hbar}[\hat{H}_{eff},\hat{\rho}]-\mathcal{\mathcal{D}}^{i}\left[\hat{\rho}\right]-\mathcal{\mathcal{D}}^{c}\left[\hat{\rho}\right]+\mathcal{M}\left[\hat{\rho}\right].\label{eq:esme}
\end{align}
The effective Hamiltonian $\hat{H}_{eff}=-\hbar\sum_{\alpha=\uparrow,\downarrow}2\Delta_{\alpha}\chi_{\alpha}\sum_{k=1}^N\left|g_{\bar{\alpha},k}\right\rangle \left\langle g_{\bar{\alpha},k}\right|$ describes the Raman-induced collective frequency shift $2\Delta_{\alpha}\chi_{\alpha}$ of the hyper-fine ground states $\left|g_{\bar{\alpha},k}\right\rangle$. Here, $k,N$ denote the individual atoms and the number of atoms, respectively, and the symbol $\overline{\alpha}$ is $\uparrow (\downarrow)$ if $\alpha=\downarrow (\uparrow)$. The abbreviations $\chi_{\uparrow(\downarrow)}=g^{2}\left|\beta_{\uparrow(\downarrow)}\right|^{2}[\Delta_{\uparrow(\downarrow)}^{2}+\gamma^{2}/4]^{-1}$ are determined by the atom-cavity mode coupling strength $g$, the field amplitudes $\beta_{\uparrow(\downarrow)}$ inside the cavity, and other parameters.  The Lindblad terms 
$\mathcal{D}^{i}\left[\hat{\rho}\right]=\gamma\sum_{\alpha=\uparrow,\downarrow}\chi_{\bar{\alpha}}\{\frac{2}{3}\sum_{k}\mathcal{D}\left[\left|g_{\alpha,k}\right\rangle \left\langle g_{\alpha,k}\right|\right]\hat{\rho}+\frac{1}{3}\sum_{k}\mathcal{D}[\left|g_{\bar{\alpha},k}\right\rangle \left\langle g_{\alpha,k}\right|]\hat{\rho}\}$ describe the Raman-induced individual dephasing (first term in the bracket), and the Raman-induced individual decay and pumping (second term in the bracket) [Fig. \ref{fig:system-theory} (b)]. 
 The Lindblad term $\mathcal{\mathcal{D}}^{c}\left[\hat{\rho}\right]=\left(4g^{2}/\kappa\right)\sum_{\alpha=\uparrow,\downarrow}\chi_{\bar{\alpha}}\mathcal{D}\left[\sum_{k}\left|g_{\alpha,k}\right\rangle \left\langle g_{\alpha,k}\right|\right]\hat{\rho}$ describes the Raman-induced collective dephasing of the atoms, where $\kappa$ is the intra-cavity photon loss rate.
The  measurement backaction of the balanced homodyne detection takes the form $\mathcal{M}\left[\hat{\rho}\right]=\left(dW/dt\right)\mathcal{H}\left[\hat{b}_{m}\right]\hat{\rho}$, and the photocurrent difference of the two photodetectors can be computed as $I(t)={\rm Re}\langle \hat{b}_m \rangle(t) + dW/dt $, which is proportional to the real part of the mean value $\langle \hat{b}_m \rangle$  but is dominated by the photon shot-noise $dW/dt$. Here, the photon annihilation operator of the measured field takes the form $\hat{b}_{m}=\sum_{\alpha=\uparrow,\downarrow}\xi_{\bar{\alpha}}\sum_{k}\left|g_{\alpha,k}\right\rangle \left\langle g_{\alpha,k}\right|$ with abbreviations $\xi_{\uparrow(\downarrow)}=\left(\beta^2_{\uparrow(\downarrow)}/\beta_{in}\right)\sqrt{\eta\kappa}\left(2g^2/\kappa\right)\left[\Delta_{\uparrow(\downarrow)}-i\gamma/2\right]^{-1}$.


We apply the collective numbers approach mentioned in the introduction section to solve Eq.~(\ref{eq:esme}), and  obtain a closed set of equation: 
\begin{equation}
\partial_t\left\langle n\right\rangle =\left(\partial_t\left\langle n\right\rangle \right)_{e}+\left(\partial_t\left\langle n\right\rangle \right)_{i}+\left(\partial_t\left\langle n\right\rangle \right)_{c}+\left(\partial_t\left\langle n\right\rangle \right)_{m}. \label{eq:col-sme}
\end{equation}
The Hamiltonian $\hat{H}_{eff}$ leads to 
\begin{equation}
\left(\partial_t\left\langle n\right\rangle \right)_{e}=-i2\sum_{\alpha=\uparrow,\downarrow}\Delta_{\bar{\alpha}}\chi_{\bar{\alpha}}\left(n_{\alpha\bar{\alpha}}-n_{\bar{\alpha}\alpha}\right)\left\langle n\right\rangle.
\end{equation} 
The dissipation of individual atoms $\mathcal{\mathcal{D}}^{i}\left[\hat{\rho}\right]$ results in
\begin{align}
 &\left(\partial_t\left\langle n\right\rangle \right)_{i}=-\gamma\sum_{\alpha=\uparrow,\downarrow}\chi_{\bar{\alpha}}\Bigl\{\Bigl[\frac{1}{2}\left(n_{\alpha\bar{\alpha}}+n_{\bar{\alpha}\alpha}\right) 
  \nonumber \\
  &+\frac{1}{3}n_{\alpha\alpha}\Bigr]\left\langle n\right\rangle -\frac{1}{3}n_{\bar{\alpha}\bar{\alpha}}\langle \begin{array}{c}
n_{\bar{\alpha}\bar{\alpha}}-1\\
n_{\alpha\alpha}+1
\end{array}\rangle \Bigr\}.
\end{align}
To ensure the compactness of the notations, here and in the following, we denote the terms on the left side of the equations by the numbers, which have changed with respect to the term on the right side of equations, e.g. $\langle \begin{array}{c}
n_{\uparrow\uparrow}-1\\
n_{\downarrow\downarrow}+1
\end{array}\rangle = \langle \begin{array}{cc}
n_{\uparrow \uparrow} -1 & n_{\uparrow \downarrow}\\
n_{\downarrow \uparrow} & n_{\downarrow \downarrow}
+1\end{array}\rangle$. The collective dephasing $\mathcal{\mathcal{D}}^{c}\left[\hat{\rho}\right]$ contributes to 
\begin{equation}
\left(\partial_t\left\langle n\right\rangle \right)_{c}=-2g^{2}/\kappa\sum_{\alpha=\uparrow,\downarrow}\chi_{\bar{\alpha}}\left(n_{\alpha\bar{\alpha}}-n_{\bar{\alpha}\alpha}\right)^{2}\left\langle n\right\rangle,
\end{equation} while the measurement backaction $\mathcal{\mathcal{M}}\left[\hat{\rho}\right]$ results in 
\begin{align}
  &\left(\partial_t\left\langle n\right\rangle \right)_{m}=-(dW/dt)\left\langle n\right\rangle  \{ (\langle \hat{b}_{m}\rangle+\langle\hat{b}_{m}^{\dagger}\rangle) \nonumber \\
  &-\sum_{\alpha=\uparrow,\downarrow}\left[\xi_{\bar{\alpha}}\left(n_{\alpha\alpha}+n_{\bar{\alpha}\alpha}\right)+\xi_{\alpha}^{*}\left(n_{\bar{\alpha}\bar{\alpha}}+n_{\alpha\bar{\alpha}}\right)\right]\}
\end{align}
with the expectation values
$ \langle \hat{b}_{m}\rangle+\langle\hat{b}_{m}^{\dagger}\rangle =\sum_{l=0}^{N}C_{N}^{l}\Bigl[\left(\xi_{\uparrow}+\xi_{\uparrow}^{*}\right)\left(N-l\right) +\left(\xi_{\downarrow}+\xi_{\downarrow}^{*}\right)l\Bigr]\langle \begin{array}{cc}
l & 0\\
0 & N-l
\end{array}\rangle
$.
Here, the binomial function is defined as $C_{n}^{m}=n!/\left[m!\left(n-m\right)!\right]$.

In order to solve Eq. \eqref{eq:col-sme}, we have to also specify the initial values $\left\langle n\right\rangle_{0}$ with the so-called  coherent spin states (CSS). These states are the product states $\left|d_{\downarrow},d_{\uparrow}\right\rangle =\prod_{k}\left|d_{\downarrow},d_{\uparrow}\right\rangle _{k}$ of individual atoms $\left|d_{\downarrow},d_{\uparrow}\right\rangle_{k}=d_{\downarrow}\left|g_{\downarrow,k}\right\rangle +d_{\uparrow}\left|g_{\uparrow,k}\right\rangle$, where the two  complex numbers can be specified with two angles $d_{\uparrow}=\sin\left(\theta/2\right)e^{i\phi}$
and $d_{\downarrow}=\cos\left(\theta/2\right)$. From these states we can construct initial density operator $\rho_{0}=\left|d_{\downarrow},d_{\uparrow}\right\rangle \left\langle d_{\downarrow},d_{\uparrow}\right|$ \cite{YZhang}, and then calculate the initial collective density matrix $\left\langle n\right\rangle _{0}=\prod_{a,b=\uparrow,\downarrow}\left(d_{a}d_{b}^{*}\right)^{n_{ab}}.$ The spin squeezing can be normally characterized by the following parameter
\cite{AS=0000F8rensen}
\begin{equation}
\xi_{n_{1}}^{2}=\frac{N\left(\Delta J_{n_{1}}\right)^{2}}{\left\langle J_{n_{2}}\right\rangle ^{2}+\left\langle J_{n_{3}}\right\rangle ^{2}}.\label{eq:squeezing-parameters}
\end{equation}
In this expression,  the collective spin operators are defined as  $\hat{J}_{x(y)}=\left(1(i)/2\right)\left(\sum_{k}\left|g_{\downarrow,k}\right\rangle \left\langle g_{\uparrow,k}\right|+(-)\sum_{k}\left|g_{\uparrow,k}\right\rangle \left\langle g_{\downarrow,k}\right|\right)$, $\hat{J}_{z}=\left(1/2\right)\left(\sum_{k}\left|g_{\uparrow,k}\right\rangle \left\langle g_{\uparrow,k}\right|-\sum_{k}\left|g_{\downarrow,k}\right\rangle \left\langle g_{\downarrow,k}\right|\right)$. The expectation values of these operators can be calculated as 
 $
 J_{x(y)}=+(-)\sum_{l=0}^{N}C_{N}^{l}l{\rm Re}({\rm Im})\langle \begin{array}{cc}
l-1 & 0\\
1 & N-l
\end{array}\rangle
$,
$
 J_{z}=(1/2)\sum_{l=0}^{N}C_{N}^{l}(2l-N)\langle \begin{array}{cc}
l & 0\\
0 & N-l
\end{array}\rangle$. Interestingly,  the former two components depend on the off-diagonal elements with  $n_{\downarrow \uparrow}=1$,  and the latter component depends  on the diagonal elements.  The squared uncertainties $(\Delta J_{n_{i}})^{2}=\langle (\hat{J}_{n_{i}})^{2}\rangle -J_{n_{i}}^{2}$ depend also on the expectation value of the squared collective
operators  
$
 \langle (\hat{J}_{x(y)})^{2}\rangle =+(-)\frac{1}{4}\sum_{l=0}^{N}C_{N}^{l}[l\left(l-1\right)\langle \begin{array}{cc}
l-2 & 0\\
2 & N-l
\end{array}\rangle 
  +(-)N\langle \begin{array}{cc}
l & 0\\
0 & N-l
\end{array}\rangle +(-)2l\left(N-l\right)\langle \begin{array}{cc}
l-1 & 1\\
1 & N-l-1
\end{array}\rangle 
  +\left(N-l\right)\left(N-l-1\right)\langle \begin{array}{cc}
l & 2\\
0 & N-l-2
\end{array}\rangle],
$
and 
$\langle (\hat{J}_{z})^{2}\rangle =\frac{1}{4}\sum_{l=0}^{N}C_{N}^{l}\left(2l-N\right)^{2}\langle \begin{array}{cc}
l & 0\\
0 & N-l
\end{array}\rangle$.  In this case, the former two quantities depend on the off-diagonal elements with either $n_{\uparrow\downarrow}=0,n_{\downarrow\uparrow}=2$  or $n_{\uparrow\downarrow}=1,n_{\downarrow\uparrow}=1$   or $n_{\uparrow\downarrow}=2,n_{\downarrow\uparrow}=0$ , and the latter one depends on the diagonal elements.

Under the off-resonant conditions,  the population of the excited states
$P_{e_{\uparrow(\downarrow),k}}$ is related to that of the ground states
$P_{g_{\downarrow(\uparrow),k}}$ as $P_{e_{\uparrow(\downarrow),k}}=\chi_{\uparrow(\downarrow)}P_{g_{\downarrow(\uparrow),k}}$. To eliminate the excited states, we require $\chi_{\uparrow(\downarrow)}\ll1$ or  $
g^{2}\left|\beta_{\uparrow(\downarrow)}\right|^{2}\ll\Delta_{\uparrow(\downarrow)}^{2}+\gamma^{2}/4$. The rates $\gamma\chi_{\uparrow(\downarrow)}$ determine the dissipation of individual atoms while the rates $N\left(4g^{2}/\kappa\right)\chi_{\uparrow(\downarrow)}$ determine the collective dephasing and the measurement backaction. Therefore, the ratio of them $N\left(4g^{2}/\kappa\right)\chi_{\uparrow(\downarrow)}/\gamma\chi_{\uparrow(\downarrow)}=N\mathcal{C}$ (single-atom cooperativity $\mathcal{C}=4g^{2}/\kappa\gamma$) determines which process dominates the system dynamics. Thus, for system with enough atoms, we may achieve $N\mathcal{C}\gg1$ such that the collective dissipation and the measurement dominate, and the  the conditional spin squeezing might be achieved.

\begin{center}
\begin{table}
\begin{centering} 
\caption{System parameters for the simulations in the main text. \label{tab:parameters}}
\begin{tabular}{|c|c|c|c|}
\hline
$\omega_{\uparrow\downarrow}$ & \hspace*{0.5cm} $2\pi\times 1.56$ GHz\hspace*{0.5cm}& $N$ & \hspace*{0.5cm} $100$\hspace*{0.5cm} \tabularnewline
\hline
$\Delta_{\uparrow}$ &\hspace*{0.5cm} $2\pi$ GHz\hspace*{0.5cm} & $\Delta_{\downarrow}$ &\hspace*{0.5cm} $2\pi$ GHz\hspace*{0.5cm} \tabularnewline
\hline
$\kappa$ &\hspace*{0.5cm} $2\pi\times 3.0$ MHz\hspace*{0.5cm} & $g$ &\hspace*{0.5cm} $2\pi\times 1.5$ MHz\hspace*{0.5cm} \tabularnewline
\hline
$\gamma$ &\hspace*{0.5cm} $2\pi\times 4.9$ MHz\hspace*{0.5cm} & $\eta$ &\hspace*{0.5cm} $0.6$\hspace*{0.5cm} \tabularnewline
\hline
$\beta_{in}$ &\hspace*{0.5cm} $120$\hspace*{0.5cm} & $\vartheta$ &\hspace*{0.5cm} $0$\hspace*{0.5cm} \tabularnewline
\hline
$\theta$ &\hspace*{0.5cm} $0.5\pi$ \hspace*{0.5cm} & $\phi$ &\hspace*{0.5cm} $0$\hspace*{0.5cm} \tabularnewline
\hline
\end{tabular}
\par\end{centering}

\end{table}
\par\end{center}

\section{Numerical Results\label{sec:results}}

Before presenting the numerical results, we discuss firstly the parameters in our simulations. We  assume that the optical cavity has a loss rate $\kappa=2\pi\times 3.0$ MHz, and couples with the atoms with a strength  $g=2\pi\times 1.5$ kHz.  The atoms have a spontaneous emission rate $\gamma=2\pi\times 4.9$ MHz, and thus the single particle cooperativity is estimated as  $\mathcal{C}=4g^{2}/\kappa\gamma\approx6.12\times10^{-7}$.
For more than  $10^{7}$ atoms, the collective cooperativity becomes  $N\mathcal{C}>1$,  the collective decay   dominates over the individual decay, and the conditional spin squeezing should work. Note that more than $10^{12}$ atoms are involved in the experiments~\citep{AEBNielsen}, and such a system might be handled with a stochastic mean-field approach \cite{YuanZhang}. To investigate the conditional spin squeezing for systems with hundreds of atoms, we increase the coupling by three orders of magnitude, $g=2\pi\times 1.5$ MHz, and obtain the single-particle cooperativity $\mathcal{C}\approx0.6$, and the collective cooperativity  $N\mathcal{C}\approx60>1$ for $N=100$. Note that the large coupling can be achieved by reducing the mode volume of the  optical cavity. 

In Tab.~\ref{tab:parameters}, we summarize the reference parameters used in the simulations, and indicate the variable parameters in the figure caption or the text. We assume that the  probe field is detuned from the atomic transitions  by the amount $\Delta=\Delta_{\uparrow}=\Delta_{\downarrow}=2\pi$ GHz,  and creates  $\left|\beta_{\uparrow}\right|^{2}=1.44\times 10^{4}$, $\left|\beta_{\downarrow}\right|^{2}=0$ photons inside the optical cavity. In this case, we  obtain  $\chi_{\uparrow}\approx3.22\times 10^{-2}$, $\chi_{\downarrow}=0$  and estimate the Raman-induced frequency shift 
$\Delta_{\uparrow}\chi_{\uparrow}\approx 2\pi\times 32.2$ MHz, $\Delta_{\downarrow}\chi_{\downarrow} = 0$ MHz, the Raman-induced individual decay $\frac{1}{3}\chi_{\downarrow}\gamma =0$ Hz  and pumping  $\frac{1}{3}\chi_{\uparrow}\gamma\approx 2\pi \times 53.6$  kHz, the Raman-induced individual  dephasing  $\frac{2}{3}\chi_{\downarrow}\gamma \approx 2\pi \times 107.1$  kHz, $\frac{2}{3}\chi_{\downarrow}\gamma =0$   and  the Raman-induced collective dephasing $\chi_{\uparrow}4g^{2}/\kappa=2\pi \times 96.4$  kHz, $\chi_{\downarrow}4g^{2}/\kappa=0$ Hz.  The frequency splitting  of the hyper-fine ground states becomes  $\omega_{\uparrow\downarrow}+\Delta_{\uparrow}\chi_{\uparrow}-\Delta_{\downarrow}\chi_{\downarrow}\approx 2\pi \times 1.56$ GHz with  $\omega_{\uparrow\downarrow}= 2\pi \times 1.53$ GHz, and the dynamics is considered  in a rotating
frame with this frequency \cite{LKThomsen,AEBNielsen}.  

In the following, we examine firstly the conditional dynamics of an ideal system in the absence of the dissipation of individual atoms, and study then the 
the influence of the probe field and the detection efficiency,  the spontaneous emission and the number of atoms.  In these simulations, we assume that the atoms are initially prepared in a CSS with $\theta=\pi/2$ and $\phi=0$, which leads to the collective spin vector along the x-axis (with $J_{x}=N/2$ and $J_{y}=J_{z}=0$) and 
$\xi_{z}^{2}=1$, i.e. standard quantum limit \cite{LKThomsen}.

\begin{figure}
\begin{centering}
\includegraphics[scale=0.25]{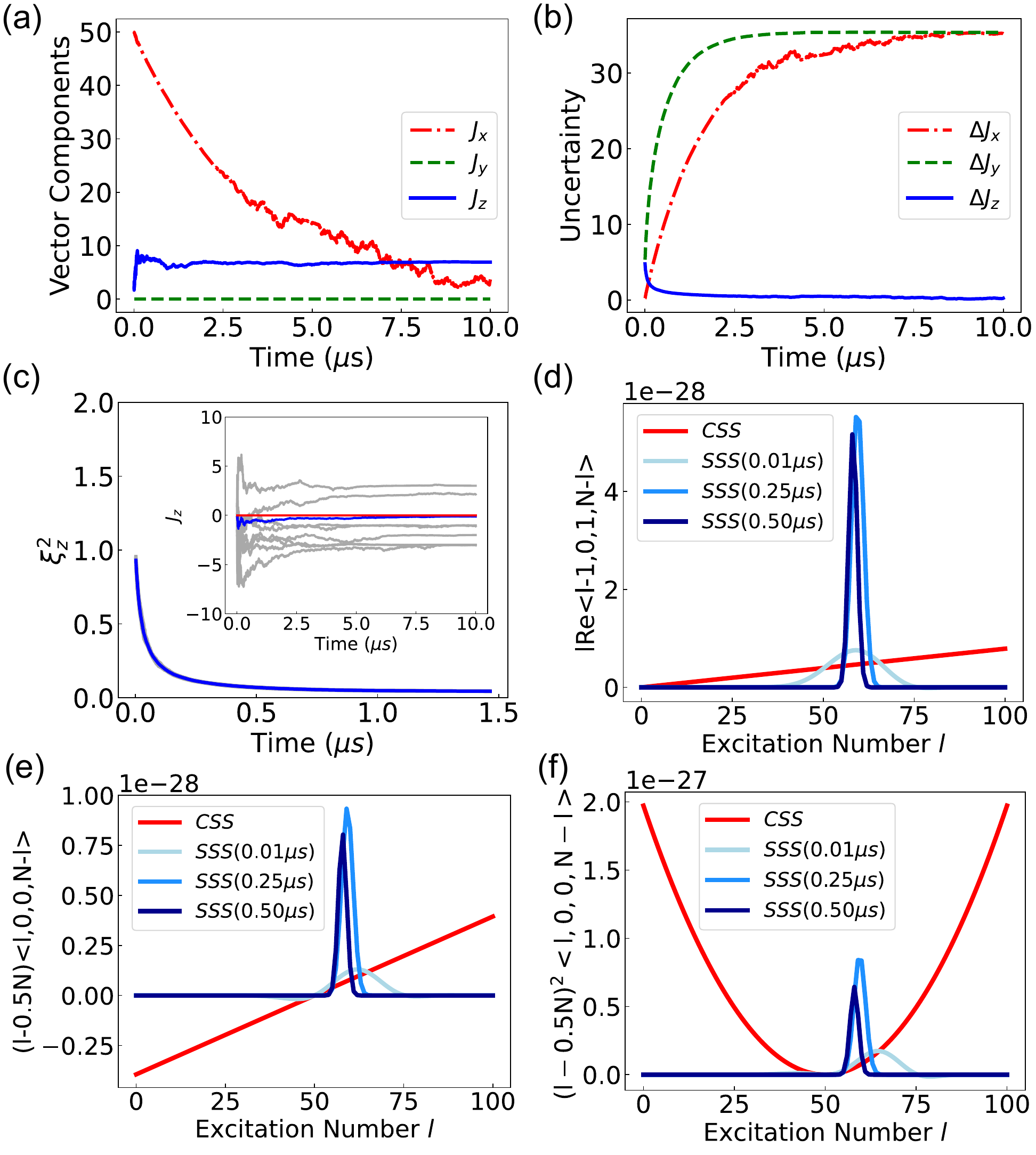}
\par\end{centering}
\caption{\label{fig:ideal} Conditional dynamics of an ideal system in the presence of homodyne detection. Panel (a) shows the three components $J_x,J_y,J_z$ of the collective spin vector, while the panel (b) shows their uncertainties $\Delta J_x,\Delta J_y,\Delta J_z$. Panel (c) shows the spin squeezing parameter $\xi_z^2$, and the $J_z$ component for different simulations (inset). Panel (d-f) show the quantities $l{\rm Re}\langle \begin{array}{cc}
l-1 & 0\\
1 & N-l
\end{array}\rangle$, $(l-N/2)\langle \begin{array}{cc}
l & 0\\
0 & N-l
\end{array}\rangle$, $(l-N/2)^2 \langle \begin{array}{cc}
l & 0\\
0 & N-l
\end{array}\rangle$ as function of the excitation quantum number $l$ for the systems in the coherent spin state (CSS,red lines) and the spin squeezed state (SSS) at time $0.01$ $\mu$s, $0.25$ $\mu$s and $0.5$ $\mu$s (light to dark blue lines). }
\end{figure}

\subsection{Conditional Dynamics of An Ideal System}

Figure~\ref{fig:ideal} shows the results for the ideal system. Figure~\ref{fig:ideal}(a) illustrates that the x-component of the collective spin vector reduces gradually from $N/2=50$ to zero in about $10$ ${\rm \mu}$s, and the z-component jumps abruptly first and remains at a value around $8$ for long time. On the contrary, the y-component remains as zero for all time. Figure~\ref{fig:ideal}(b) shows that the uncertainties of the x- and y-component increase firstly from zero and $5$, and approach steadily the constant around $35$, signaling the anti-squeezing, while the uncertainty of the z-component decreases from $5$ and approaches zero for long time, indicating the squeezing. Figure~\ref{fig:ideal}(c) shows more precisely that the spin squeezing parameter starts from unity, i.e. the standard quantum limit, and reduces dramatically, i.e. the spin squeezing, and approaches gradually to zero for long time. By carrying out many simulations, we find that  all the results are similar except that the z-component might have different signs for different simulations [inset of Fig.~\ref{fig:ideal}(c)]. 

To gain insights into the formation of the spin squeezing, we further examine the states of the atomic ensemble by checking the different elements of the collective density matrix [Fig.~\ref{fig:ideal}(d-f)]. As explained in the paragraph below Eq.~\eqref{eq:squeezing-parameters}, all the components of the collective spin vector are determined by a common factor $C_{N}^{l}$ besides of the collective density elements. Since the binomial function $C_{N}^{l}$ shows a broad peak for $l=N/2$ [Fig.~\ref{fig:Jz}(f)], the elements with $l$ around $N/2$ should contribute most to the vector components. Since the x,z-component $J_x,J_z$ of the collective spin vector and the uncertainty $\Delta J_z$ of the z-component are related to $l{\rm Re}\langle \begin{array}{cc}
l-1 & 0\\
1 & N-l
\end{array}\rangle$, $(l-N/2)\langle \begin{array}{cc}
l & 0\\
0 & N-l
\end{array}\rangle$, $(l-N/2)^2\langle \begin{array}{cc}
l & 0\\
0 & N-l
\end{array}\rangle$ beside of the common factor, respectively, we have examined the distribution of these term as function of excitation quantum number $l$ in Fig.~\ref{fig:ideal}(d-f). For the system in the CSS, the collective matrix elements are constant, and thus the former two quantities increase linearly with $l$ and the latter quantity shows a parabolic dependence on $l$. In the contrast, for the system in the SSS, all the quantities show peaks around $l\approx 60$ for short probing time $0.01$ $\mu$s, and these peaks become much sharper and shift to smaller $l$ for much longer probing time. Obviously, the narrowing of these peaks is correlated with the spin squeezing, and the shift of these peaks is correlated with the change of $J_z$. Since these peaks resemble the Gaussian distributions, they justify the Gaussian state formalism as often used in the studies of spin squeezing~\citep{BLMadsen}. In addition, for longer probe time, the first and third quantity reduce, and the second quantity remains unchanged. These dynamics are consistent with the evolution of $J_x,\Delta J_z, J_z$ , respectively.

\begin{figure}
\begin{centering}
\includegraphics[scale=0.17]{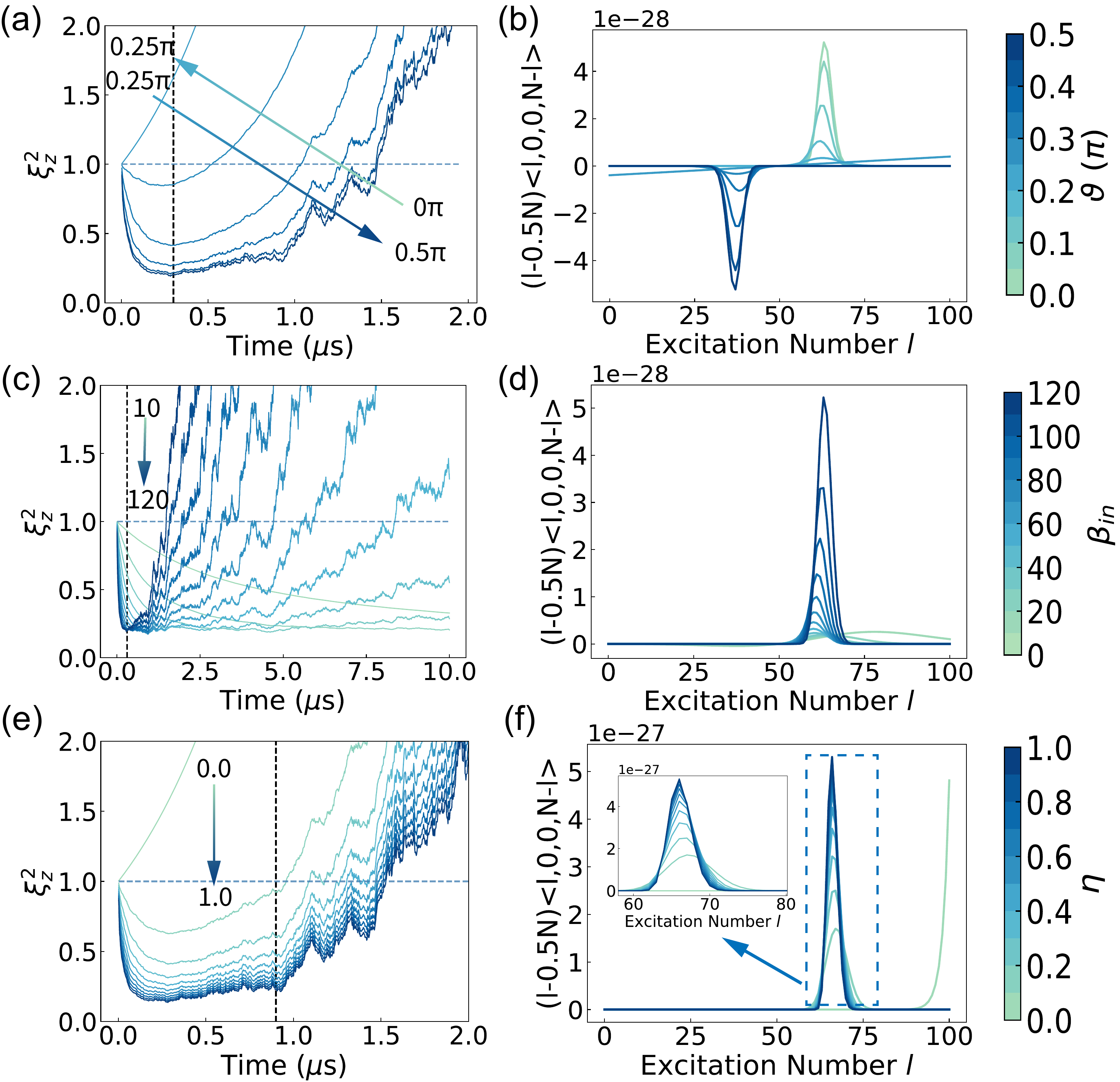}
\par\end{centering}
\caption{\label{fig:probe} Influence of  the probe field direction as  parameterized by the angle $\vartheta$ (a,b), the probe field amplitude $\beta_{in}$  (c,d) and the detection efficiency of photondetectors $\eta$ (e,f) on the spin squeezing parameter $\xi_z^2$ (a,c,e) and the quantity $(l-N/2)\langle \begin{array}{cc}
l & 0\\
0 & N-l
\end{array}\rangle$ for SSS at different time. In the panels (a,c,e), the horizontal dashed lines indicate  the standard quantum limit with $\xi_z^2=1$, and the vertical dashed lines indicate the time considered in the panel (b,d,f).  The same list of the random number ${dW}$ are used here for better visualisation, and the influence of different random numbers is presented in Fig.~\ref{fig:random} of the Appendix. For other parameters see Tab.~\ref{tab:parameters} and  the text. }
\end{figure}

\subsection{Influence of Probe Field and Detection Efficiency}

Figure \ref{fig:probe} shows the influence of the  polarization (a,b) and amplitude (c,d) of the probe field as well as the detection efficiency  (e,f) on the spin squeezing parameter $\xi_{z}^{2}$  (a,c,e) and the quantity $(l-N/2)\langle \begin{array}{cc}
l & 0\\
0 & N-l
\end{array}\rangle$  (b,d,f), where the related vector z-component  $J_z$ is shown in Fig.~\ref{fig:Jz} of the Appendix. To provide a complete picture, in Fig.~\ref{fig:components}, we have also shown all the vector components $J_x,J_y$, $J_z$ and their uncertainties $\Delta J_x,\Delta J_y, \Delta J_z$ for the reference situation. Here, we assume that the projection of the probe-field polarization on the two cavity modes  $\beta_{\uparrow}=\cos\vartheta\beta_{in}$ and $\beta_{\downarrow}=\sin\vartheta\beta_{in}$ can be determined by an angle $\vartheta$. To understand these results, we  rewrite the measurement backaction as $\mathcal{M}\left[\hat{\rho}\right]=\left(dW/dt\right) (\xi_\downarrow -\xi_\uparrow)\mathcal{H}[\hat{J}_z]\hat{\rho}$,  and the collective dephasing of the atoms  as $\mathcal{\mathcal{D}}^{c}\left[\hat{\rho}\right]=\left(4g^{2}/\kappa\right) (\chi_\downarrow + \chi_\uparrow) \mathcal{D}\left[\hat{J}_z\right]\hat{\rho} $.  Using the expressions for $\beta_{\uparrow},\beta_{\downarrow}$, we obtain  $\xi_\downarrow -\xi_\uparrow \propto \eta |\beta_{in}|^2{\rm cos}(2\vartheta)$, and $\chi_\downarrow + \chi_\uparrow \propto  |\beta_{in}|^2$. Furthermore, we have also the Raman-induced individual decay rate $\frac{1}{3}\chi_\downarrow\gamma\propto \gamma |\beta_{in}|^2 \sin^2\vartheta$  and pumping rate $\frac{1}{3}\chi_\uparrow\gamma\propto \gamma |\beta_{in}|^2 \cos^2\vartheta$, as well as the Raman-induced individual dephasing rate $\frac{2}{3}\gamma(\chi_\downarrow + \chi_\uparrow)\propto \gamma |\beta_{in}|^2$.   From these expressions, we note that the collective and individual dephasing are angle-independent,  the measurement backaction,  the individual decay and pumping depend on the angle $\vartheta$  in different ways, and all the processes are proportional to the intensity of the probe field $ |\beta_{in}|^2$. 

According to the above expressions,  the parameter $\xi_\downarrow -\xi_\uparrow$, which characterizes the strength of the QND measurement and the measurement backaction,  is maximal for  $\vartheta = 0,0.5\pi$, where the probe field is parallel with the polarization of one of the cavity modes. This parameter reduces slightly for $\vartheta$ slightly departing from these angles, and becomes strictly zero for  $\vartheta=0.25\pi$, where the probe field has equal projection along the polarization of the two cavity modes. Thus, we expect that the spin-squeezing is optimal  in the former case, and  becomes weaker in the later case, as confirmed by  Fig.  \ref{fig:probe} (a). More precisely, for $\vartheta=0,0.5\pi$, the squeezing parameter $\xi_{z}^{2}$  drops dramatically from {one (i.e. the standard quantum limit) in short time,   then decays slowly to a minimal value, and raises finally above one in long time. The anti-squeezing in long time is caused by the reduced length of the collective spin vector, which can be further attributed to the Raman-induced collective and individual dephasing (Fig.~\ref{fig:components} of the Appendix).

Besides, the Raman-induced pumping  $\frac{1}{3}\gamma\chi_\uparrow$  dominates over the Raman-induced decay  $\frac{1}{3}\gamma\chi_\downarrow$  for the angle $\vartheta<0.25\pi$, and the opposite case is valid for the angle  $\vartheta>0.25\pi$. Thus, we expect that the Gaussian-like distribution of the collective density matrix elements is shifted to larger $l>N/2$ and smaller $l<N/2$ in the former and latter case, and should have the largest value for $\theta = 0,\pi/2$,  as shown in Fig.~\ref{fig:probe} (b). Figure.~\ref{fig:probe} (b) shows that there is a slight shift in the  Gaussian-like distribution of the collective density matrix elements and the absolute values of the peaks become smaller and larger under the former and the latter case, and have the largest value for $\theta = 0,\pi/2$. Since the two processes are equally strong for the particular angle  $\vartheta=0.25\pi$,  we expect that the atomic ensemble is not squeezed, and the distribution of the collective density matrix elements becomes linear. Accompanying with these results, the vector z-component $J_z$ becomes positive and negative in the former ($\vartheta<0.25\pi$) and latter case ($\vartheta>0.25\pi$), and does not change so much from the initial value for the particular angle ($\vartheta=0.25\pi$) [Fig.~\ref{fig:Jz}(a)]. Note that  the Raman-induced pumping, decay and the measurement backaction affect the vector component $J_z$, but not affect the length of the vector projection in the equator plane (Fig.~\ref{fig:components} of the Appendix).

Furthermore, the parameter $\xi_{\downarrow}-\xi_{\uparrow}$ is linearly proportional to the square of the probe field $|\beta_{in}|^2$. Thus, we expect both the better spin squeezing and the strong measurement backaction. Indeed, as shown in Fig.  \ref{fig:probe}  (c,d),  for  increasing  $\beta_{in}$,  the  initial falling of the squeezing parameter  $\xi_{z}^{2}$   is more dramatic, and the Gaussian-like distribution of the collective density matrix elements becomes more pronounced and sharper. Correspondingly, the raise of $J_z$ is also much more severe [Fig.~\ref{fig:Jz}(b)]. At the same time, the Raman-induced collective and individual dephasing are also proportional to the the square of the probe field $|\beta_{in}|^2$. Thus, as this parameter increases, the length of the collective spin vector projection in the equator plane reduces more dramatically (Fig.~\ref{fig:probecomp} of the Appendix). Since the spin squeezing parameter $\xi_z^2$ is inversely proportional to the square of this length, the spin squeezing becomes compromised, and the later raise of $\xi_{z}^{2}$  is much steeper and occurs also much earlier.

Since the measurement backaction is directly proportional to the detection efficiency $\eta$, its influence on the system dynamics becomes more pronounced as this parameter increases, leading to the better spin squeezing. Indeed, with increasing  $\eta$, the squeezing parameter $\xi_z^2$ decays much faster and the minimal squeezing reduces [Fig.~\ref{fig:probe}(e)],  and the Gaussian-like distribution of the collective density matrix elements becomes more significant and narrower [Fig.~\ref{fig:probe}(f)]. Accordingly, the the vector component $J_z$ becomes much larger [Fig.~\ref{fig:Jz}]. Furthermore, we confirm that the Raman-induced  collective  and individual dephasing are responsible for the reduced length of the collective vector projection and an raise in $\xi_{z}^2$ in longer time by examining systems with different polarization angles (Fig.~\ref{fig:components} of the Appendix).

In the above simulations, the same list of the random numbers ${dW}$ are used for better visualisation. However,  in Fig.~\ref{fig:random} of the Appendix, we present also the results for different random numbers, and find that the spin squeezing parameter $\xi_z^2$, the population distribution (thus the collective spin vector  z-component $J_z$) change similarly except that the small random features change from simulation to simulation. By averaging the results for many simulations, both the parameters become smooth, and the $J_z$ curve approaches the one in the absence of the measurement.

\begin{figure}
\begin{centering}
\includegraphics[scale=0.25]{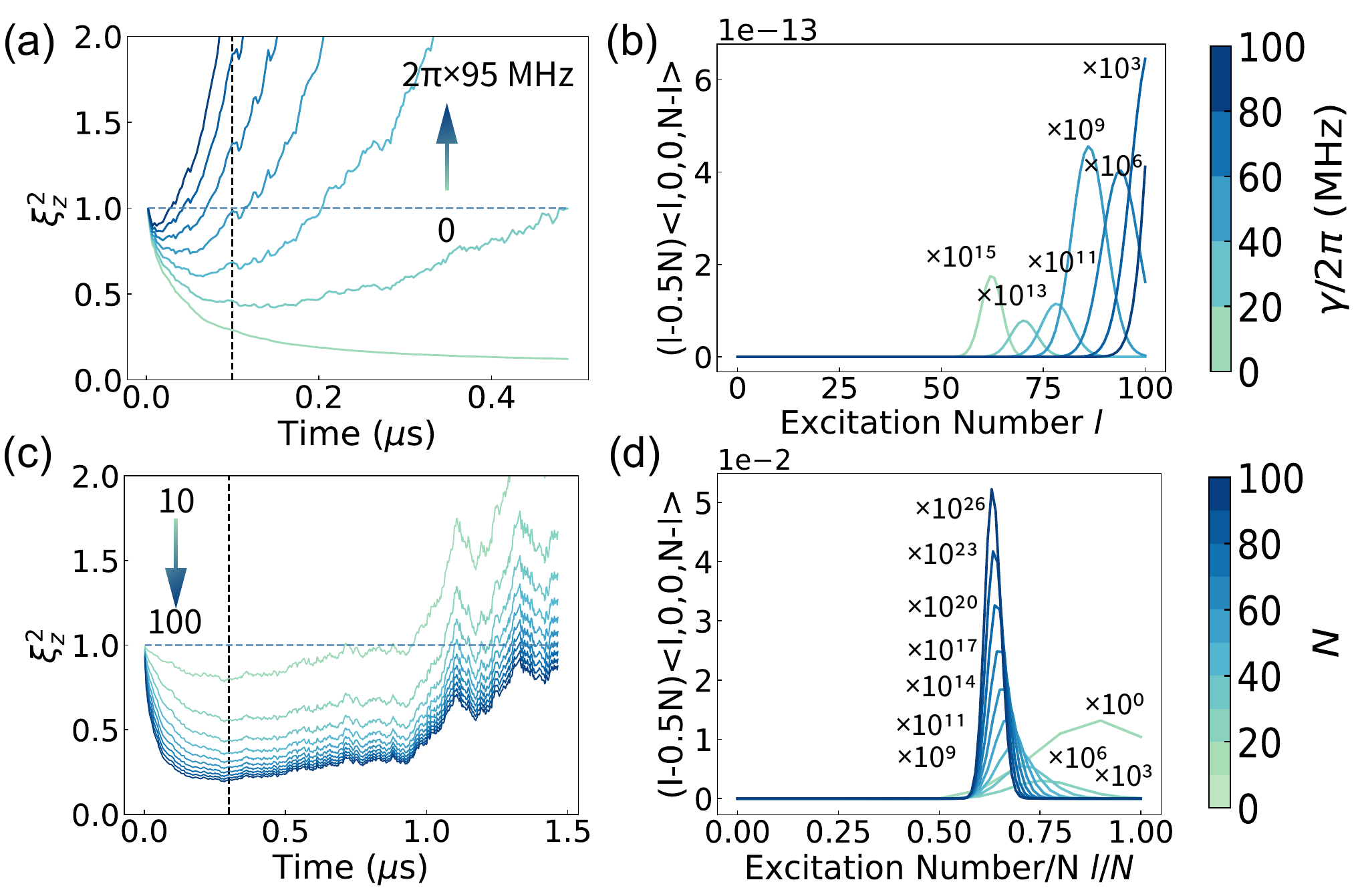}
\par\end{centering}
\caption{\label{fig:decay} Influence of the spontaneous emission rate $\gamma$ (a,b) and  the number of atoms $N$ (c,d) on the spin squeezing parameter $\xi_z^2$ (a,c) and the quantity $(l-N/2)\langle \begin{array}{cc}
l & 0\\
0 & N-l
\end{array}\rangle$ for given time (b,d), as marked by the vertical dashed line in the panel (a,c).  In the panels (a,b), one hundred atoms $N=100$ are considered. In the panels (c,d), the numbers near the curves indicate the amplification factors. For better visualisation, the same list of the random number ${dW}$ are used in the simulations.  For other parameters see Tab.~\ref{tab:parameters} and  the text.
}
\end{figure}

\subsection{Influence of Spontaneous Emission and Atom Number}

In the above simulations, we have included the the Raman-induced pumping, decay and dephasing of the atoms, which are caused by the unavoidable spontaneous emission of the individual atoms. Even in this case, the spin squeezing can be achieved by the probe field with proper polarization, amplitude and the photon-detector with high efficiency. In Fig. \ref{fig:decay}, we investigate the influence of the spontaneous emission rate $\gamma$ in more details, and also demonstrate that this influence can be compensated partially by adding more atoms into the system. 

In Fig. \ref{fig:decay}(a,b), we vary $\gamma$ arbitrarily from zero to $2\pi \times 95$ MHz, where the later value is much larger than the realistic value $2\pi \times 4.9$ MHz. As the spontaneous emission rate $\gamma$ increases,  $\xi_z^2$ reaches the minimal value and becomes larger than one at much earlier time, and the minimal value approaches steadily to one. This means that the time window for the spin squeezing shrinks, and the available degree of squeezing becomes limited. At the same time, the Gaussian-like distribution of the collective density matrix elements shifts to larger $l$,  becomes broader, and increases also by orders of magnitude. Correspondingly, $J_z$ is no longer fixed at some constant values, but increases steadily to much larger value [Fig.~\ref{fig:Jz}(d)]. As explained before, the former behavior can be attributed to the Raman-induced collective and individual dephasing of atoms, and the latter is caused by the domination of the Raman-induced pumping over the Raman-induced decay.

In Fig.~\ref{fig:decay}(c,d), we keep the spontaneous emission rate as the value in the Tab.~\ref{tab:parameters}, but increase the number of atoms $N$ in the simulations. For $N=10$, the time window for the spin squeezing $\xi_z^2<1$ is about $0.8$ ${\rm \mu s}$, and the minimal squeezing parameter is only about $0.8$.  For much larger $N$, the time window of the spin squeezing increases steadily, and the minimal squeezing becomes also much smaller.  Accompanying with this, the Gaussian-like distribution of the collective density matrix elements shifts to $l$ near to $N/2$, becomes much sharper, reduces by orders of magnitude. Here, the orders of magnitude reduction is compensated by the opposite increase of the common factor $C_N^l$. As a result, $J_z$ becomes much larger [Fig.~\ref{fig:Jz}(e)], and the ratio $J_z/N$ reduces actually, indicating a smaller measurement backaction per atom. In summary, indeed, the detrimental effect of the spontaneous emission can be mitigated by working with more atoms, and the better spin squeezing can be achieved in this case.

\section{Conclusions \label{sec:conclusions}}

In summary,  we have presented an exact approach to solve the stochastic master equation based on the collective density matrix, and studied the conditional spin squeezing of hundreds of atoms caused by the homodyne detection of the light-shift in Raman processes. Our calculations show that the Gaussian-like distribution of the collective density matrix elements is formed for the spin squeezed state, and the optimal conditional spin squeezing is achieved for the probe field with proper polarization and moderate strength, the photondetectors with larger detection efficiency, the system with smaller spontaneous emission rate, and larger number of atoms. 

To observe the conditional spin squeezing with hundreds of atoms, we have increased the single particle-cavity coupling, which might be achieved by reducing the mode volume. Otherwise, we have to consider systems with more atoms. To this end,  we  can apply the Monte Carlo wave-function approach to explore only portion of the Hilbert space \citep{KM=0000F8lmer,KJacobs}, or use the stochastic mean-field approach \cite{YuanZhang} to consider the mean values instead of the density matrix.  In any case, the exact approach presented here can play a very important role in gauging these  approaches, or in exploring the entanglement effects beyond these approaches~\citep{ANegretti,SMassar}. Furthermore, the current approach can be also  extended straightforwardly to account for the optical cavity and the multiple-level atoms, which can be used to explore the conditional spin squeezing under the strong coupling conditions or the influence of the higher atomic levels.

\begin{acknowledgments}
ZhiQing Zhang carried out the numerical calculations under the supervision of Yuan Zhang who developed the theory and the numerical programs. They contribute equally to the work. All authors contributed to the analyses and the writing of the manuscript. This work is supported by the National Key R\&D Program of China under Grant No. 2021YFA1400900, the National Natural Science Foundation of China under Grants No. 12004344, 12174347, 12074232, 12125406, 62027816, U21A2070, and the Cross-disciplinary Innovative Research Group Project of Henan Province No. 232300421004, as well as the Carlsberg Foundation through the Semper Ardens grant "QCooL". 
\end{acknowledgments}

\appendix

\section{Stochastic Master Equation for Total System \label{sec:sme}}
In this Appendix, we present the stochastic master equation for the  
 total density operator $\hat{\rho}$  of the system shown in Fig. \ref{fig:system-theory} (a) in the rotation frame of the probe field 
\begin{equation}
\partial_t\hat{\rho}=-\frac{i}{\hbar} [\hat{H}_{a}+\hat{H}_{a-p},\rho ]-\mathcal{\mathcal{D}}^{a}\left[\hat{\rho}\right]-\mathcal{\mathcal{D}}^{f}\left[\hat{\rho}\right] + \mathcal{M}\left[\hat{\rho}\right].\label{eq:total-master-equation}
\end{equation}
The Hamiltonian $\hat{H}_{a}=\hbar\Delta_{\uparrow}\sum_{k=1}^{N}\left|e_{\uparrow,k}\right\rangle \left\langle e_{\uparrow,k}\right|+\hbar\Delta_{\downarrow}\sum_{k}\left|e_{\downarrow,k}\right\rangle \left\langle e_{\downarrow,k}\right|$ describes the ensemble of atoms with two hyper-fine excited states $\left|e_{\uparrow,k}\right\rangle$ and $\left|e_{\downarrow,k}\right\rangle$. $k,N$ denotes the individual atom and the number of atoms, respectively. The atoms have also two hyper-fine ground states $\left|g_{\uparrow,k}\right\rangle $ and $\left|g_{\downarrow,k}\right\rangle$. Here, $\Delta_{\uparrow}$ ($\Delta_{\downarrow}$) is the frequency difference of the transitions $g_{\uparrow,k}\to e_{\downarrow,k}$ ($g_{\downarrow,k}\to e_{\uparrow,k}$) to the probe field. The atom-probe field  coupling is given by $\hat{H}_{a-p}=\hbar g[\sum_{\alpha=\uparrow,\downarrow}\left(i\beta_{\alpha}+\hat{c}_{\alpha}\right)\sum_{k}\left|e_{\alpha,k}\right\rangle \left\langle g_{\bar{\alpha},k}\right|+ \sum_{\alpha=\uparrow,\downarrow}\left(-i\beta_{\alpha}^*+\hat{c}_{\alpha}^\dagger \right)\sum_{k}\left|g_{\bar{\alpha},k}\right\rangle \left\langle e_{\alpha,k}\right|]$ with the atom-cavity mode coupling coefficient $g$, the amplitude of the probe field $\beta_{\uparrow},\beta_{\downarrow}$, the creation operator $\hat{c}_{\uparrow}^\dagger$ and annihilation operator $\hat{c}_{\uparrow}$ of scattered photons. Notice that $\bar{\alpha}=\uparrow,\downarrow$ if $\alpha=\downarrow,\uparrow$. We account for the spontaneous emission of the atomic excited states with the Lindblad terms $\mathcal{D}^{a}\left[\hat{\rho}\right]=\gamma\sum_{\alpha=\uparrow,\downarrow}(\frac{2}{3}\sum_{k=1}^N\mathcal{D}\left[\left|g_{\bar{\alpha},k}\right\rangle \left\langle e_{\alpha,k}\right|\right]\hat{\rho}+\frac{1}{3}\sum_{k=1}^N\mathcal{D}\left[\left|g_{\alpha,k}\right\rangle \left\langle e_{\alpha,k}\right|\right]\hat{\rho})$ and the loss of the intra-cavity scattered photons $\mathcal{\mathcal{D}}^{f}\left[\hat{\rho}\right]=\kappa\left(\mathcal{D}\left[\hat{c}_{\uparrow}\right]\hat{\rho}+\mathcal{D}\left[\hat{c}_{\downarrow}\right]\hat{\rho}\right).$

In the following, we analyze the measurement backaction of the homodyne detection $\mathcal{M}\left[\hat{\rho}\right]$. We can decompose the probe field as two components with polarization along $\mathbf{e}_{\uparrow}$ and $\mathbf{e}_{\downarrow}$ directions $\mathbf{e}_{in}\beta_{in}=\mathbf{e}_{\uparrow}\mathbf{\beta}_{\uparrow}+\mathbf{e}_{\downarrow}\beta_{\downarrow}$. Obviously, we have $\beta_{\uparrow}=\mathbf{e}_{in}\cdot\mathbf{e}_{\uparrow}\beta_{in}$ and $\beta_{\downarrow}=\mathbf{e}_{in}\cdot\mathbf{e}_{\downarrow}\beta_{in}$. The field out of the cavity can be written as $\mathbf{e}_{out} \hat{b}_{out}=\mathbf{e}_{in}\beta_{in}+\sqrt{\eta\kappa}\left(\mathbf{e}_{\uparrow}\hat{c}_{\uparrow}+\mathbf{e}_{\downarrow}\hat{c}_{\downarrow}\right)$. In the balanced homodyne detection, we measure the phase difference of the probe field, which is described by $\mathbf{e}_{m}\hat{b}_{m}=\sqrt{\eta\kappa}\left(\mathbf{e}_{\uparrow}\hat{c}_{\uparrow}+\mathbf{e}_{\downarrow}\hat{c}_{\downarrow}\right)$ with $\hat{b}_{m}=\hat{b}_{out}-\beta_{in}$ [Fig. \ref{fig:system-theory}
(a)]. Assuming $\mathbf{e}_{m}=\mathbf{e}_{in}=\mathbf{e}_{out}$ we have $\hat{b}_{m}=\sqrt{\eta\kappa}\left(\frac{\beta_{\uparrow}}{\beta_{in}}\hat{c}_{\uparrow}+\frac{\beta_{\downarrow}}{\beta_{in}}\hat{c}_{\downarrow}\right)$, where  $\eta$  describes the  efficiency of the photon-detectors.   Finally, we get the measurement backaction \cite{HMWiseman,LKThomsen}  $\mathcal{M}\left[\hat{\rho}\right]=\left(dW/dt\right)\mathcal{H}[\hat{b}_{m}]\hat{\rho}$  of the homodyne detection with the photon-shot noise, where the Wiener increment  $dW$  follows a  normal distribution with mean $E[dW]=0$ and variance $dW^2=dt$ (with $dt$ as the simulation time-step). The photoncurrent difference of the two detectors can be computed as $I(t)={\rm Re}\langle \hat{b}_m \rangle(t) + dW/dt $, which is proportional to the real part of the mean value $\langle \hat{b}_m \rangle$  but is dominated by the photon shot-noise $dW/dt$

To achieve the effective stochastic master equation~\eqref{eq:esme}, we eliminate further adiabatically the two excited hyper-fine states. To do so, we simply carry out the following replacements $\left|e_{\uparrow(\downarrow),k}\right\rangle \to g\beta_{\uparrow(\downarrow)}^{*}\left[\Delta_{\uparrow(\downarrow)}+i\gamma/2\right]^{-1}\left|g_{\downarrow(\uparrow),k}\right\rangle $
and $\hat{c}_{\uparrow(\downarrow)}\to2\left(g^2/\kappa\right)\beta_{\uparrow(\downarrow)}\left[\Delta_{\uparrow(\downarrow)}-i\gamma/2\right]^{-1}\sum_{k}\left|g_{\downarrow(\uparrow),k}\right\rangle \left\langle g_{\downarrow(\uparrow),k}\right|$ and ignore the terms proportional to $\Delta_{\uparrow(\downarrow)}$ and $g$ in Eq. \eqref{eq:total-master-equation}. At the same time, the measured
field operator can be written as $\hat{b}_{m}=\sum_{\alpha=\uparrow,\downarrow}\xi_{\bar{\alpha}}\sum_{k}\left|g_{\alpha,k}\right\rangle \left\langle g_{\alpha,k}\right|$ with abbreviations $\xi_{\uparrow(\downarrow)}=\left(\beta^2_{\uparrow(\downarrow)}/\beta_{in}\right)\sqrt{\eta\kappa}\left(2g^2/\kappa\right)\left[\Delta_{\uparrow(\downarrow)}-i\gamma/2\right]^{-1}.$  Notice that in the expression $\xi_{\uparrow(\downarrow)}\xi_{\uparrow(\downarrow)}^{*}=\left|\beta_{\uparrow(\downarrow)}/\beta_{in}\right|^{2}\eta\left(4g^{2}\chi_{\uparrow(\downarrow)}/\kappa\right)$, the terms in the brackets is the collective dephasing rate in $\mathcal{\mathcal{D}}^{c}\left[\hat{\rho}\right]$.

\section{Program Codes to Solve Stochastic Collective Density Matrix Equations}

To simulate the system with as many atoms as possible, we have utilized the parallel computation technique enabled by NVIDIA graph card to speed up the numerical evolution of Eq.~\eqref{eq:col-sme}. The corresponding codes are provided in the supplementary material~\citep{SuppMat}. There, the "parameters.cuh" section specifies the system parameters as micro values. The "functions.cu" section defines the ancillary functions, and the "functions.cuh" section include the declarations of these functions.  The "random.py" section is used to generate a set of random numbers to describe the Wiener process. The "dynamics.cu" section defines the main function to represent, and solve Eq.~\eqref{eq:col-sme}, and save the results.

In these codes, the most crucial thing is how to define  and access the collective density matrix elements $\left\langle n\right\rangle \equiv\left\langle \begin{array}{cc}
n_{\uparrow \uparrow} & n_{\uparrow \downarrow}\\
n_{\downarrow \uparrow} & n_{\downarrow \downarrow}
\end{array}\right\rangle $.  We define one list with  $C_{N+3}^3$  elements to present  $\left\langle n\right\rangle$. To access these elements,  we utilize three nested loops. In the out-most loop, we let  the number $n_{\downarrow\downarrow}$ take the value from zero to $N$. In the second loop, we let the number $n_{\downarrow\uparrow}$ take the value from zero to $N-n_{\downarrow\downarrow}$. In the inner-most loop, we let the number $n_{\uparrow\downarrow}$ vary from zero to  $N-n_{\downarrow\downarrow}-n_{\downarrow\uparrow}$ and calculate the number $n_{\uparrow\uparrow}=N-n_{\downarrow\downarrow}-n_{\downarrow\uparrow}-n_{\uparrow\downarrow}$. In this way, we can generate  all the set of numbers $\{n_{\uparrow\uparrow},n_{\uparrow\downarrow},n_{\downarrow\uparrow},n_{\downarrow\downarrow}\}$ so that they satisfy the conditions $0\le n_{ab}\le N$ and $\sum_{a,b=\uparrow,\downarrow} n_{ab} =N$. In addition, we define a variable $ind$  and initialize it as $-f(N,3)-f(N,2)-f(N,1)$, and increase it in the inner-most loop. In this way, we can associate each set  to one index through the equation 
\begin{align}
ind &= \sum_{i=0}^{n_{\downarrow\downarrow}} f(N-i,3) + \sum_{i=0}^{n_{\downarrow\uparrow}} f(N-n_{\downarrow\downarrow}-i,2) \nonumber \\ 
& + \sum_{i=0}^{n_{\uparrow\downarrow}} f(N-n_{\downarrow\downarrow}-n_{\downarrow\uparrow}-i,1) +1. \label{eq:relation}
\end{align}
Here, we have defined the function $f(n,k)=C^{k-1}_{n+k-1}$. 

During the numerical evolution of $\left\langle n\right\rangle $, we have to utilize Eq.~\eqref{eq:relation} to calculate the index of given element. However, by analyzing Eq.~\eqref{eq:col-sme}, we find that the index for this element can be evaluated in a simple way. For example, the index $ind_{1}$ for   $\langle n_{\uparrow\uparrow},n_{\uparrow\downarrow},n_{\downarrow\uparrow}+1,n_{\downarrow\downarrow}-1\rangle$  can be calculated from the index $ind_{0}$  of  
$\langle n_{\uparrow\uparrow},n_{\uparrow\downarrow},n_{\downarrow\uparrow},n_{\downarrow\downarrow}\rangle$ according to the relation $ind_{1} = ind_{0} + F(N-n_{\downarrow\downarrow},1)$ . Here, we have defined the function $F(n,m)=f(n,m+1)-f(n,m+2)$.

\section{Extra Results}

In this Appendix, we provide extra numerical results to complement the discussions in the main text.

\begin{figure}
\begin{centering}
\includegraphics[scale=0.25]{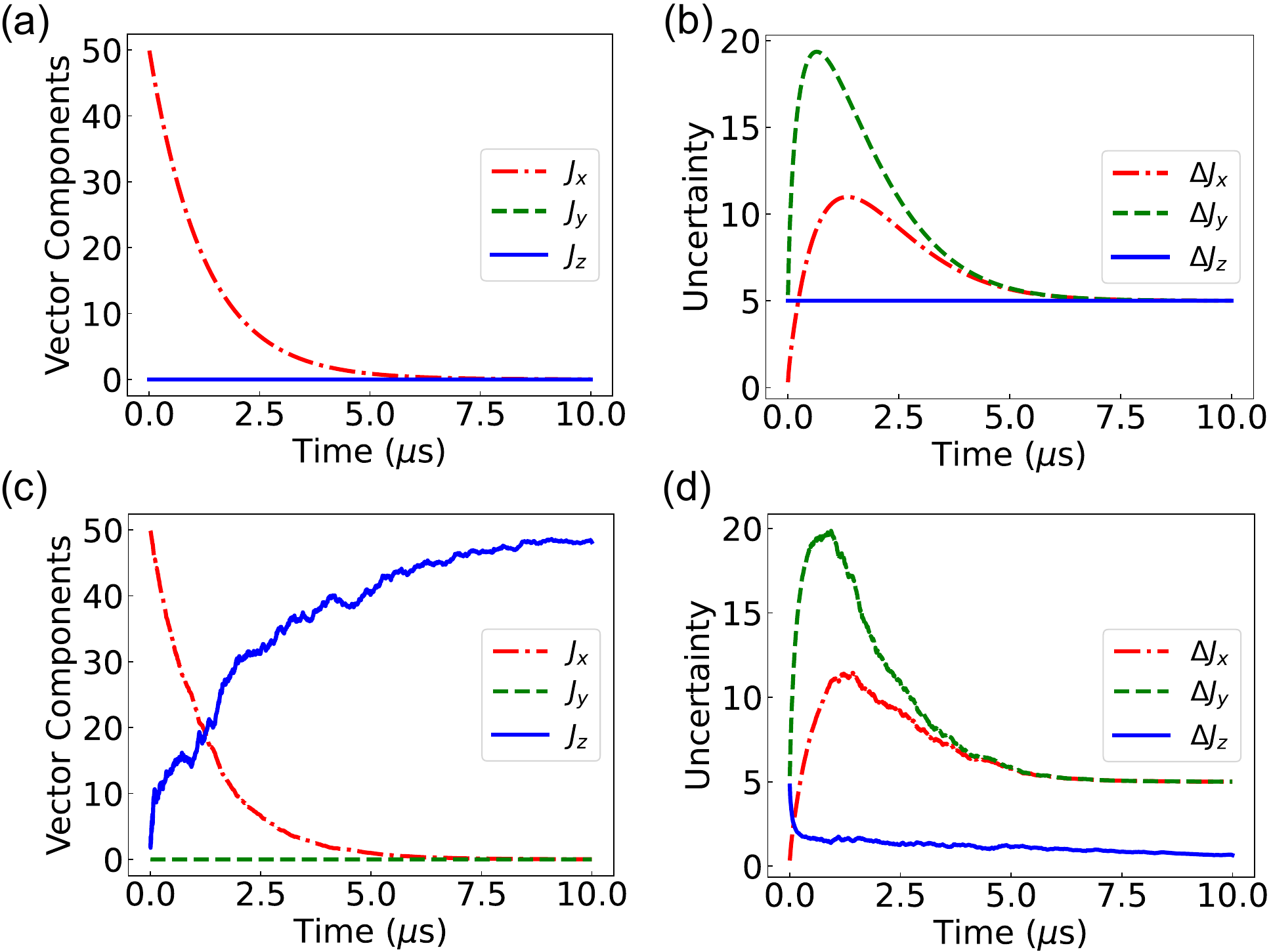}
\par\end{centering}
\caption{\label{fig:components}  Evolution of the collective spin vector components $J_x,J_y,J_z$ (a,c) and their uncertainties $\Delta J_x, \Delta J_y, \Delta J_z$ (b,d) for the system with $\theta = 0.25\pi$ (a,b) and $\theta = 0\pi$ (c,d). For other parameters, see Tab.~\ref{tab:parameters}.}
\end{figure}

\subsection{Spin Vector Components and Uncertainties}

In the main text, we have focused on the evolution of the spin squeezing parameter $\xi_z^2$, and the z-component of the collective spin vector $J_z$. To provide a complete picture, in Fig.~\ref{fig:components},  we show the evolution of the collective spin vector components $J_x,J_y,J_z$ (a,c) and their uncertainties  $\Delta J_x, \Delta J_y, \Delta J_z$ (b,d). Furthermore, we examine also the cases with angle $\theta = 0.25 \pi$ (a,b) and $\theta = 0$ (c,d), where the QND measurement is effectively absent and present, respectively. 

In the case without the QND measurement, the vector components $J_y,J_z$ are always zero, and the component $J_x$ reduces exponentially, which is due to the presence of the Raman-induced collective dephasing $\chi_\downarrow + \chi_\uparrow \propto  |\beta_{in}|^2$ and individual dephasing $\gamma(\chi_\downarrow + \chi_\uparrow) \propto  \gamma|\beta_{in}|^2$. At the same time, the uncertainty $\Delta J_z$ remains as 5, while the other two $\Delta J_x,\Delta J_y$ increase first and then decrease gradually to the steady state value of 5.

In the case with the QND measurement, the vector components behave similarly, except that the z-component $J_z$ increases steadily and approaches the maximal value $50$ for longer time. At the same time, the component uncertainties behave also similarly except that the uncertainty of the z-component $J_z$ reduces fast in short time, but decreases slowly at long time. Since the spin squeezing parameter $\xi_z^2$ is directly proportional to $\Delta J_z^2$ but inversely to $J_x^2 + J_y^2$, we except that this parameter drops fast below one in short time, but raises over one in long time, as demonstrated in the main text.

\begin{figure}
\begin{centering}

\includegraphics[scale=0.50]{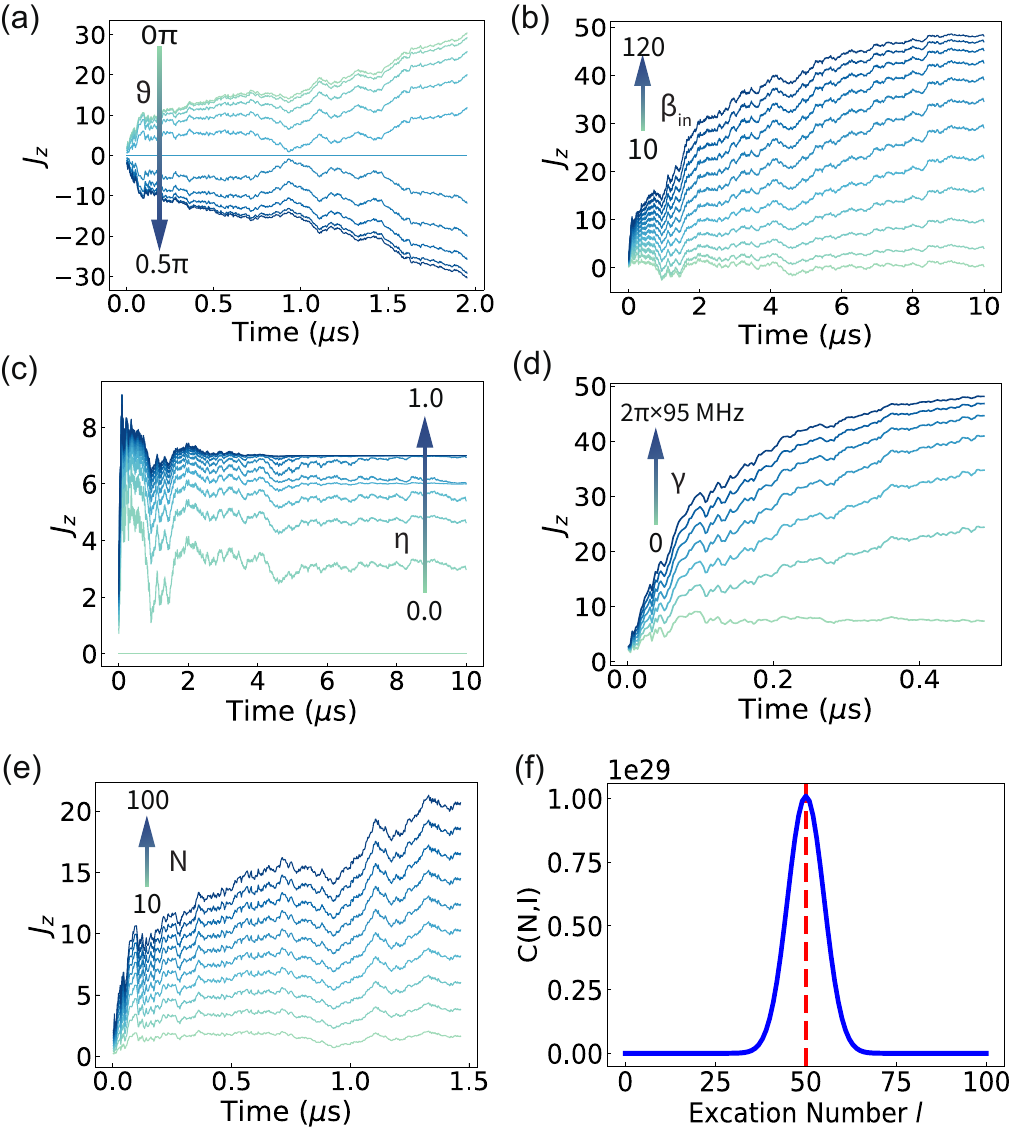}
\par\end{centering}
\caption{\label{fig:Jz} Evolution of collective spin vector z-component $J_z$ for different probe field direction as parameterized by the angle $\vartheta$ (a), probe field amplitude $\beta_{in}$ (b), detection efficiency of photodetectors (c), spontaneous emission rate $\gamma$ (d), and the number of atoms $N$ (e), respectively. In panel (c), the spontaneous emission rate is assumed as $\gamma = 0$ for a better visualisation. Panel(f) shows the variation of the binomial function $C_{N}^{l}$ with $l$. For other parameters see Tab.~\ref{tab:parameters}. }
\end{figure}

\subsection{Collective Spin Vector z-Component and Binomial Function}

In the Fig.~\ref{fig:probe} and Fig .~\ref{fig:decay} of the main text, we demonstrate the influence of different parameters on spin squeezing parameters $\xi^{2}_{z}$ and the quantity $(l-N/2)\langle \begin{array}{cc}
l & 0\\
0 & N-l
\end{array}\rangle$. In Fig .~\ref{fig:Jz}, we show the evolution of the collective spin vector z-component $J_z$ for different probe field direction as parameterized by the angle $\vartheta$ (a), probe field amplitude $\beta_{in}$ (b), detection efficiency of photodetectors $\eta$ (c), spontaneous emission rate $\gamma$ (d), and  number of atoms $N$ (e), respectively. Figure~\ref{fig:Jz}(a) shows that in general the absolute value of $J_z$ increases with time, and  $J_z$ changes from positive to negative when $\vartheta$ becomes larger than $0.25\pi$.  Figure~\ref{fig:Jz}(b) shows that as  $\beta_{in}$ increase, $J_z$ increases more faster, and the saturated value becomes much larger.  Figure~\ref{fig:Jz}(c) shows that as $\eta$ increase, $J_z$ departs more from zero.  Figure~\ref{fig:Jz}(d) shows that  $J_z$ acquires finite value for $\eta=0$, increases gradually for finite $\eta$, and becomes much larger for larger $\eta$. 
Figure~\ref{fig:Jz}(e) shows that as the number of atoms increases, the vector component $J_z$ gradually increases. Figure~\ref{fig:Jz}(f) shows the variation of the binomial function $C_{N}^{l}$ with the excitation $l$ when $N=100$, and $C_{N}^{l}$ follows a Gaussian distribution and has a peak at $l=N/2$.

\begin{figure}
\begin{centering}

\includegraphics[scale=0.25]{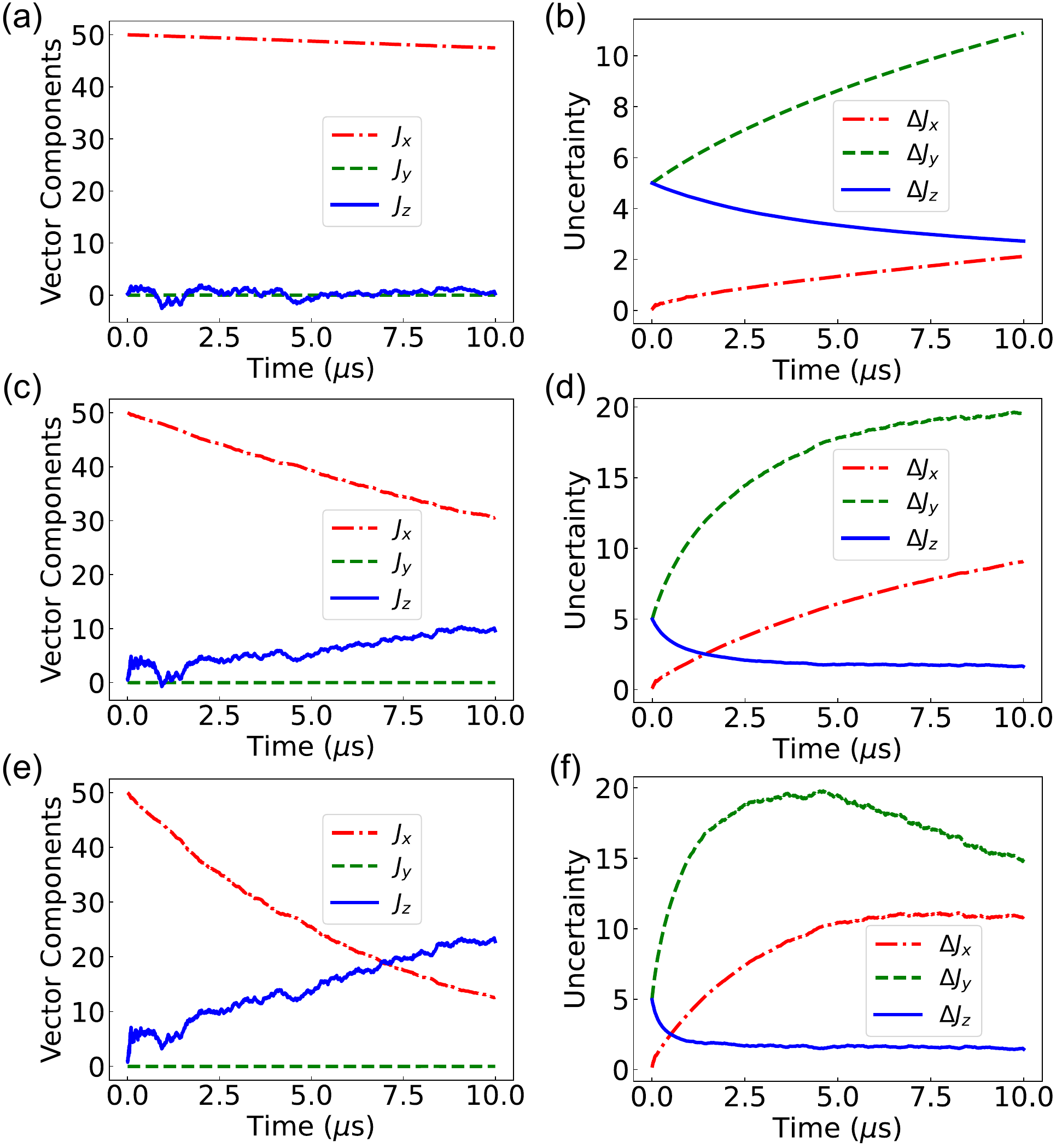}
\par\end{centering}
\caption{\label{fig:probecomp}   Evolution of the collective spin vector components $J_x, J_y, J_z$ (a,c,e) and their uncertainty  $\Delta J_x, \Delta J_y, \Delta J_z$  (b,d,f) for the probe field with the amplitude $\beta_{in}=10$ (a,b), $\beta_{in}=30$ (c,d), $\beta_{in}=50$ (e,f), respectively. For other parameters see Tab.~\ref{tab:parameters}.}
\end{figure}

\subsection{Influence of the Probe Field Amplitude}

In Fig.~\ref{fig:probe} of the main text, we discuss the influence of the probe field amplitude on the spin squeezing. In Fig.~\ref{fig:probecomp}, we provide more information by examining the collective components $J_x,J_y,J_z$ (a,c,e) and their uncertainties  $\Delta J_x, \Delta J_y, \Delta J_z$  (b,d,f)  for the probe field with different amplitude  $\beta_{in}$. We see that as $\beta_{in}$ increases, the component $J_x$ reduces much faster, and $J_z$ departs more from zero value. The former is due to the increased Raman-induced collective dephasing and individual dephasing, and the latter is caused by the increased measurement backaction and Raman-induced individual pumping. Accompanying with these changes, the uncertainty $\Delta J_z$ drops much faster, and the saturated value becomes much smaller. Since the spin squeezing parameter $\xi_z^2$ is  proportional directly to $\Delta J_z^2$ but inversely to $J_x^2 + J_y^2$, this parameter should drop much faster, but raise much earlier for increasing $\beta_{in}$, as demonstrated in Fig.~\ref{fig:probe}(c). Note that the uncertainties $\Delta J_x,\Delta J_y$ increase with time, indicating the anti-squeezing.




\begin{figure}
\begin{centering}
\includegraphics[scale=0.25]{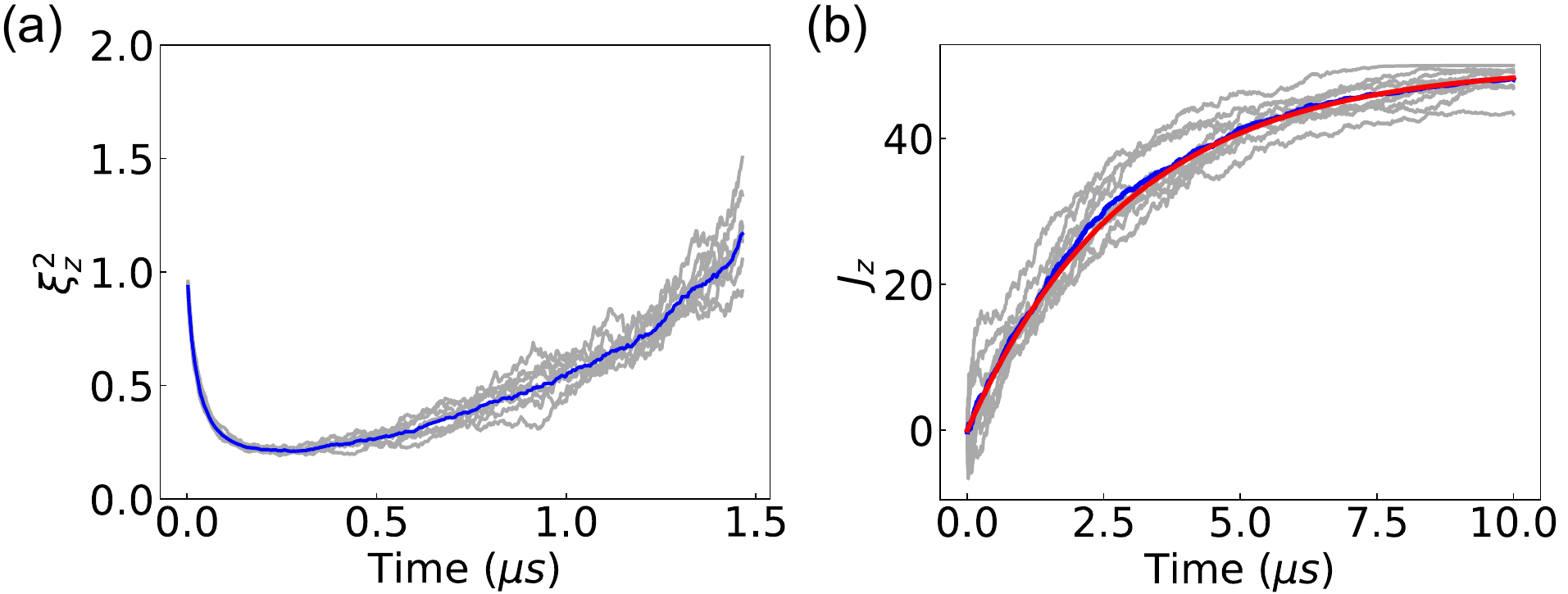}
\par\end{centering}
\caption{\label{fig:random}  Influence of different random numbers (representing the photon shot-noise) on the spin squeezing dynamics with the spontaneous emission rate $\gamma=2\pi\times 4.9$ MHz 
 MHz (a,b). The gray curves show the different simulations, the blue solid line shows the average, and the red solid line shows the results in the absence of measurements.}
\end{figure}

\subsection{Influence of Random Number}

In the main text, to better illustrate the influence of various parameters on the spin squeezing dynamics, we have utilized the same list of random numbers in all the simulations,  representing the same photon shot-noise of the photondetectors. However, in reality, the photon shot-noise noise might vary from experiment to experiment, despite that many realizations of the shot-noise follows some statistical law. Thus, in Fig.~\ref{fig:random}, we present the results for the simulations with different lists of random numbers. As we see, the spin squeezing parameter $\xi_z^2$ (a) and the z-component of the collective spin vector $J_z$ (b) evolve similarly  under eight different sets of random numbers. Although each of the curves fluctuates due to different noise, their average gives relatively smooth curves (solid blue lines), and the averaged $J_z$ curve is close to the one in the absence of measurement (solid red line).

\end{document}